\documentclass[journal,draftcls,onecolumn,12pt,twoside]{IEEEtranTCOM}

\usepackage{cite}
\usepackage{bm}
\usepackage{amsmath,amssymb,amsfonts}
\usepackage{algorithmic}
\usepackage{graphicx}
\usepackage{amsthm}
\usepackage{float}
\usepackage{balance}
\usepackage{color} 
\usepackage{algorithm}
\usepackage{subfigure}

\usepackage{setspace}
\begin{document}

\title{\LARGE A Splitting-Detection Joint-Decision Receiver for Ultrasonic Intra-Body Communications}

\author{Qianqian~Wang,~\IEEEmembership{Student Member,~IEEE}, Quansheng~Guan,~\IEEEmembership{Senior Member,~IEEE},\\
	Julian~Cheng,~\IEEEmembership{Senior Member,~IEEE}, and Fei~Ji,~\IEEEmembership{Member,~IEEE}
	
	\thanks{Qianqian Wang, Quansheng Guan, and Fei Ji are with the School of Electronic and Information Engineering, South China University of Technology, Guangzhou 510640, China (Corresponding author: Quansheng~Guan; e-mail: eeqshguan@scut.edu.cn).}
	\thanks{Julian Cheng is with the School of Engineering, The University of British Columbia, Kelowna, BC V1V 1V7, Canada.}}
\maketitle
\vspace{-2cm}
\begin{abstract}
	Ultrasonic intra-body communication (IBC) is a promising enabling technology for future healthcare applications, due to low attenuation and medical safety of ultrasonic waves for the human body. \textcolor{black}{A splitting receiver, referred to as the splitting-detection separate-decision (SDSD) receiver, is introduced for ultrasonic pulse-based IBCs, and SDSD can significantly improve bit-error rate (BER) performance over the traditional coherent-detection (CD) and energy detection (ED) receivers. To overcome the high complexity and improve the BER performance of SDSD, a splitting-detection joint-decision (SDJD) receiver is proposed.} The core idea of SDJD is to split the received signal into two steams that can be separately processed by CD and ED, and then summed up as joint decision variables to achieve diversity combining. \textcolor{black}{The theoretical channel capacity and BER of the SDSD and SDJD are derived for $M$-ary pulse position modulation ($M$-PPM) and PPM with spreading codes. {The derivation takes into account the channel noise, intra-body channel fading, and channel estimation error.} Simulation results verify the theoretical analysis and show that both SDSD and SDJD can achieve higher channel capacity and lower BER than the CD and ED receivers with perfect channel estimation, while SDJD can achieve the lowest BER with imperfect channel estimation.}
\end{abstract}


\section{Introduction} \label{sect: introduction}
\textcolor{black}{The advance of ultra-miniaturized implantable devices has enabled applications for future healthcare systems, such as continuous and accurate monitoring of the physiological and pathological conditions of soft tissues, nanorobots for targeted drug delivery, etc. \cite{Uspower,b1,b2,b5}. These applications demand effective and safe intra-body communications (IBCs). Due to high absorption and radiation in biomedical tissues, the widely used radio-frequency (RF) waves are unsuitable for IBCs.} Compared with RF waves, ultrasonic waves have much lower absorption and less interference. Thus, the ultrasonic wave is a promising medium to connect medical implants~\cite{e-health}.

Ultrasonic IBCs have attracted increasing research attention. The propagation characteristics of ultrasonic wave have shown  it is a viable medium for IBCs \cite{Y.D.medium}. The trade-offs among ultrasonic transmission frequency, transmission power, and bandwidth in IBCs have slao been investigated \cite{Challenge}. Many advanced modulation and multiplexing techniques have been applied to IBCs, e.g., quadrature amplitude modulation \cite{b9,b10}, orthogonal frequency division multiplexing \cite{b11,b12}, and multiple-input multiple-output systems \cite{b13}. \textcolor{black}{However, these techniques rely on sending continuous ultrasonic waves, which result in power-hungry devices and may even cause heat trauma for the human body  \cite{Challenge}.} Since clinical treatment and diagnosis have adopted ultrasonic pulses for many years \cite{e-health,b14}, ultrasonic wideband (UsWB) systems including time-hopping UsWB (TH-UsWB) \cite{UsWBMAC,UsWBrate} and direct-sequence UsWB (DS-UsWB) \cite{DS-USWB}, and ultrasonic index modulation (UsIM) \cite{UsIM,UsIM1} all use low duty-cycle short pulses to overcome the multipath effect and to alleviate the impact of thermal and mechanical effects. 

All of the above research has focused on modulation schemes for the transmitter in IBCs. Besides the transmitter, the receiver also determines the reliability of a communication system. There are two types of IBC receivers: coherent-detection (CD) receiver and non-coherent detection receiver (e.g., energy detection (ED) receiver) \cite{DigiCom}. The CD receiver detects the amplitude and phase of the received signal, while the ED receiver detects the energy. \textcolor{black}{However, the ultrasonic pulse signal space contains the degrees of freedom (DoFs) in amplitude, phase, and energy. Either the CD or ED receiver exploits only partial dimensions of the received signal space, which is called the DoFs in a channel \cite{Fundof}. Thus, the under-exploration of DoFs in the receiver will incur loss in signal-to-noise ratio (SNR) for the received signal.}

It has been shown that the optimal detector is a linear combination of coherent and non-coherent detectors \cite{DigiComFad}. \textcolor{black}{To achieve the optimal reception, the receiver should adopt both CD and ED to detect all DoFs \cite{Fundof,DigiComFad}. A recently proposed splitting receiver adopts independent CD and ED receivers to process received signals simultaneously, and such a receiver generates  two independent decision vectors that can be used to estimate the received RF signals \cite{R.Z.2,SDSD2017,SDSD2020}.} This splitting-detection separate-decision (SDSD) receiver can achieve higher rate and lower bit-error rate (BER) than the CD or ED receiver in high SNR regimes. \textcolor{black}{Inspired by the low BER and high data rate of the SDSD receiver, we first introduce an SDSD receiver for ultrasonic pulse-based IBCs. This SDSD architecture is theoretically realizable for IBC because existing ultrasonic power and data transfer systems are equipped with both CD and energy harvesting circuits\cite{b23,U-verse}.}

 
\textcolor{black}{This paper shows that the SDSD receiver can be used to improve the BER performance of either CD or ED receiver in IBCs, under perfect channel estimation. However, the SDSD receiver requires to implement both the CD and ED receivers, which lead to complicated hardware design for implanted devices. Additionally, the maximum-likelihood based minimum mean-square error (ML-MMSE) algorithm in the RF-based SDSD receiver \cite{SDSD2017} requires exact channel estimation and complex multiplication operations for high-order modulations. The high hardware and computational complexities make the SDSD receiver unsuitable for IBCs, which have strict requirements on energy consumption and implant size \cite{Challenge}. More importantly, it is challenging to achieve perfect intra-body channel estimation in practice, and the BER performance of SDSD can deteriorate significantly with imperfect channel estimation.}

\textcolor{black}{To tackle the aforementioned problems in the SDSD receiver, in a preliminary work, we proposed a splitting-detection joint-decision (SDJD) receiver for ultrasonic IBCs to lower the complexity of the SDSD receiver \cite{SDJD}. We have numerically compared the performance of the SDSD and SDJD receivers in the case of perfect channel estimation. In this paper, we consider a more practical channel  with channel estimation errors, and provide a comprehensive in-depth analysis and insights into the proposed SDSD and SDJD receivers. The main contributions of this paper are summarized as follows:}
\begin{itemize}
	\item \textcolor{black}{A novel SDJD receiver is proposed for ultrasonic pulse-based IBCs.
	The SDJD receiver first splits the received signal into two streams for the CD and ED, which only contain the detection phases. Then, the detected signals in SDJD are combined to jointly estimate the received ultrasonic pulses. Different from SDSD, SDJD does not require the CD and ED receivers to generate separate CD and ED signals as independent decision variables. Benefiting from the diversity combining of CD and ED signals, the SDJD receiver is superior to the SDSD receiver in terms of lower BER and lower complexity.}

	\item \textcolor{black}{We analyze the channel capacity and BER of SDSD and SDJD receivers in intra-bady fading channels. The specific receiver design, signal processing and analysis are different from the previous studies \cite{SDSD2017,SDSD2020} as the ultrasonic pulses are fundamentally different from the continuous RF waves. The optimal splitting ratios that achieve the lowest BER for SDSD and SDJD receivers are also analyzed by considering the effect of channel fading. Both theoretical and simulation results show that the SDJD receiver has lower BER than SDSD with imperfect channel estimation.}
	
\end{itemize}

\textcolor{black}{The remainder of this paper is organized as follows. In Section II, the SDSD receiver is first applied  to IBCs, and then the SDJD receiver is proposed. Section III analyzes the system performance. The simulation results are explored in Section IV, and conclusions are drawn in Section V.}

\vspace{-0.3cm}
\section{Proposed SDJD for IBCs} \label{sect: Received Signal}
In this section, a low-complexity SDJD receiver is proposed for ultrasonic IBCs. The signal models of the transmitted and received signals and channel model of IBCs are also presented. 

\subsection{Structure of SDJD}

\begin{figure}[!t]
	\centering{\includegraphics[width=4in]{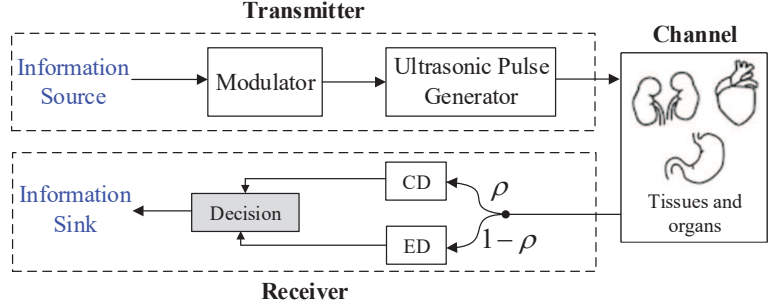}}
	\caption{An ultrasonic pulse-based IBC system using the splitting receiver.}
	\label{System1}
\end{figure}

Figure \ref{System1} illustrates an ultrasonic pulse-based IBC system using the splitting receiver. The system consists of three main parts: transmitter, intra-body channel, and receiver. The information bits are modulated to generate ultrasonic pulses. The transmitted signal is then propagated over the intra-body channel. The ubiquitous tissues and organs in the human body will cause the reflection and refraction of the transmitted pulse, leading to multipath and fading effects. The SDSD and SDJD structures have to split the received signal and detect the signal using both CD and ED to exploit all the DoFs of the signal. The difference between SDSD and SDJD exists in the decision phase. The details of the splitting receiver structure are presented in the following.

\subsubsection{Splitting the received signal}
\begin{figure}[!t]
	\centering
	\subfigure[]{
		\label{CD}
		\includegraphics[width=2.6in]{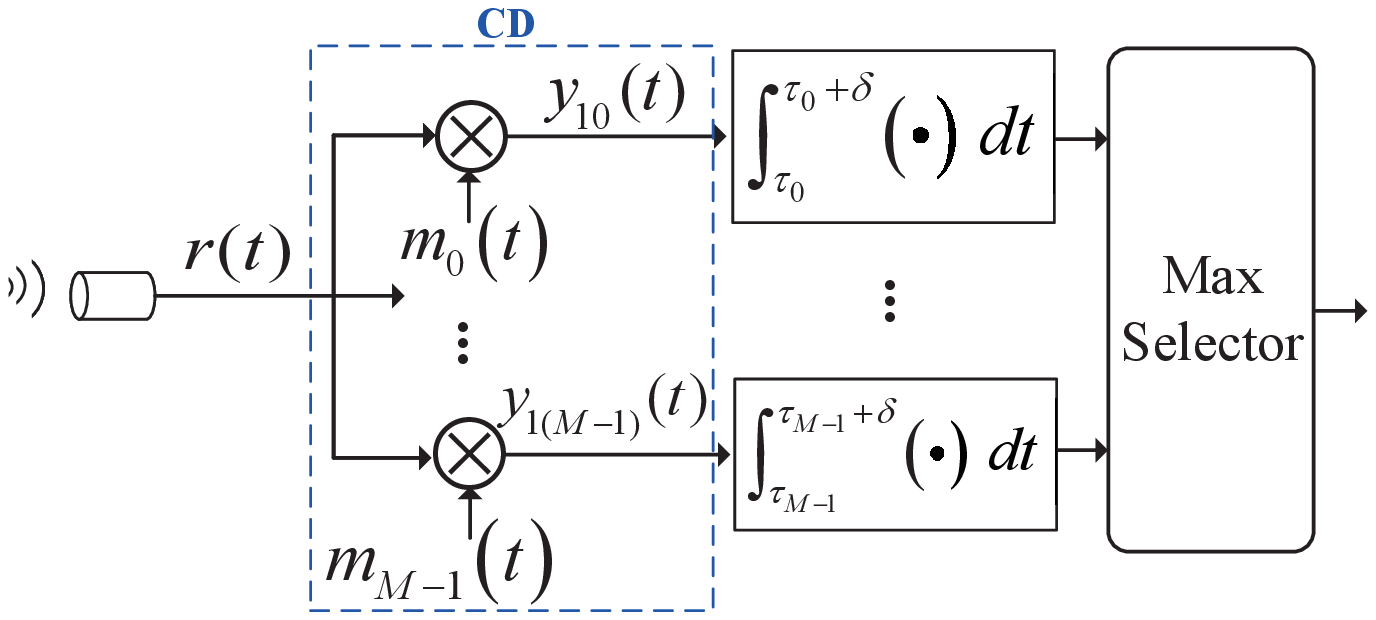}}\quad
	\subfigure[]{
		\label{ED}
		\includegraphics[width=2.9in]{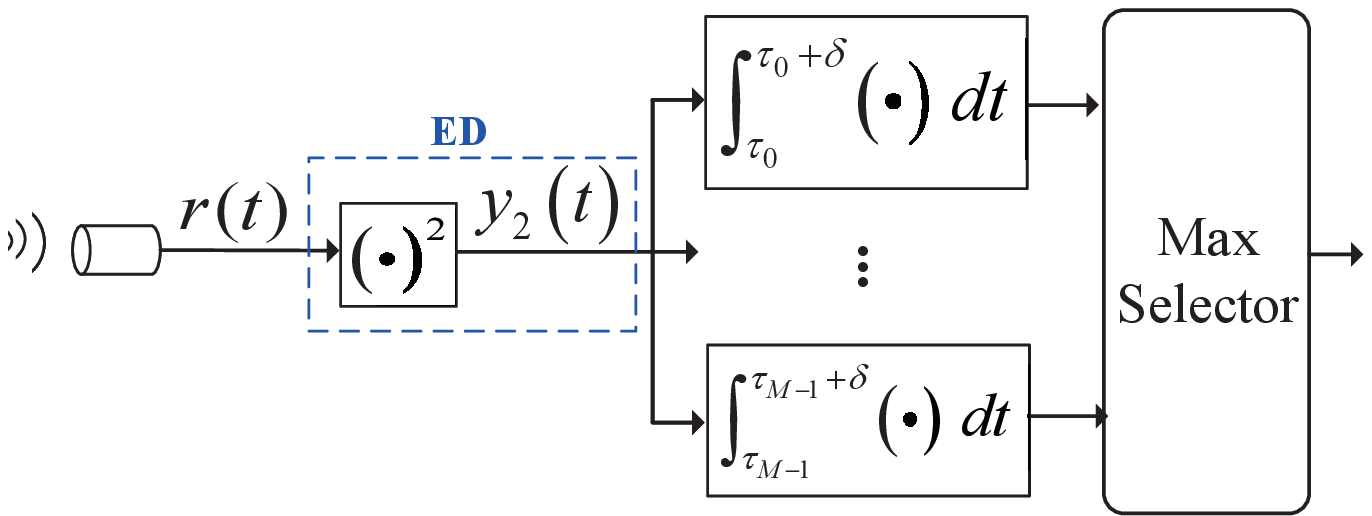}}\vspace{-0.3cm}\\
	\caption{Structures of the (a) CD receiver and (b) ED receiver.}
	\label{CD-ED}
	\vspace{-0.3cm}
\end{figure}

Traditional receivers, as shown in Fig. \ref{CD-ED}, adopt either CD or ED to generate decision variables, which are used to recover the transmitted signal. Compared with the receivers having single detection, Fig. \ref{System1} shows that the splitting detection uses both CD and ED to detect the received signal. Specifically, the received signal is first split into two streams having the power splitting ratio $\rho$ and $1-\rho$ by time switching or a power splitting device \cite{R.Z.2,R.Z.1}. The two streams then are processed by CD and ED to generate decision variables. Note that the splitting ratio $\rho=0$ means that the received signal is only detected by ED and $\rho=1$ means that the signal is only detected by CD \footnote{To distinguish the detector and the receiver, CD or ED in this paper only contains the detection phase, while the CD or ED receiver includes both the detection and decision phases.}. According to different decision criteria, the specific receiver designs include SDSD and SDJD, which will be illustrated next.

\begin{figure}[!t]
	\centering
	\subfigure[]{
		\label{SDSD}
		\includegraphics[width=3.8in]{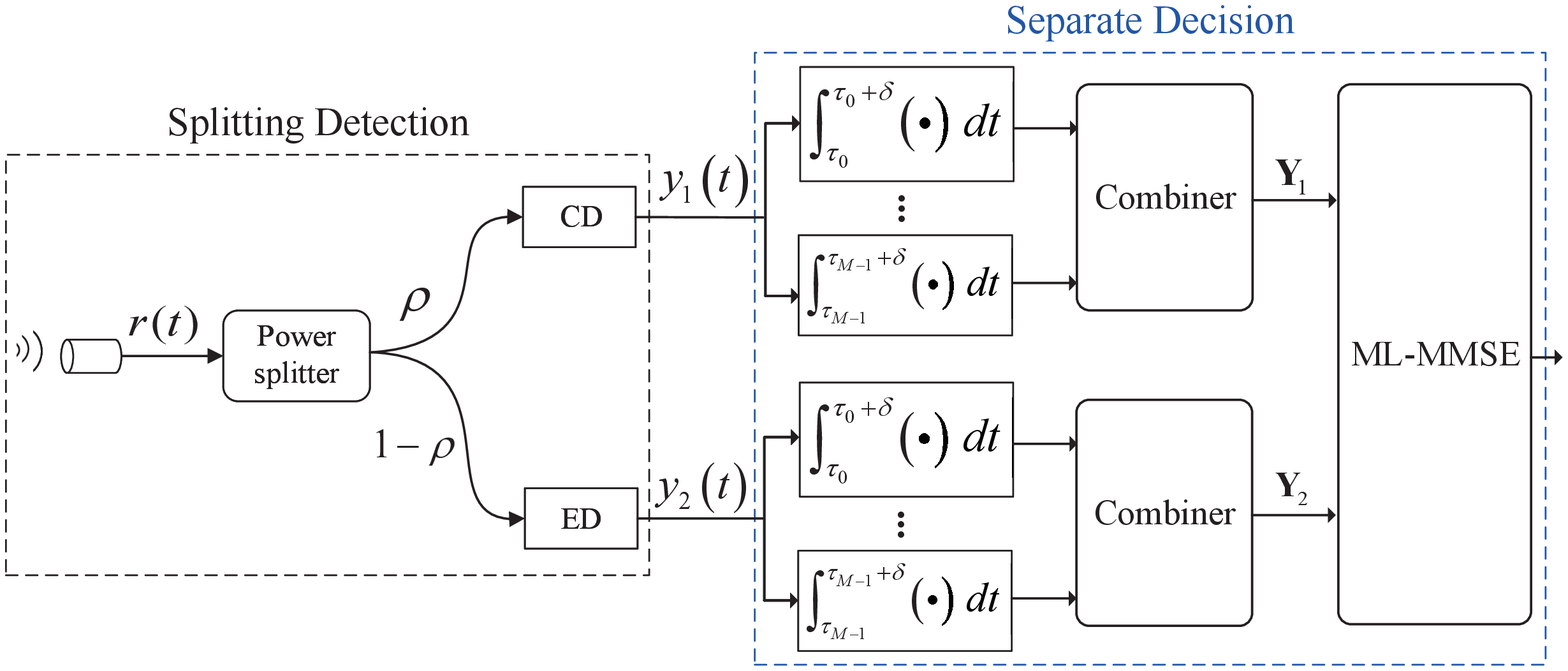}}
	\subfigure[]{
		\label{SDJD}
		\includegraphics[width=3.8in]{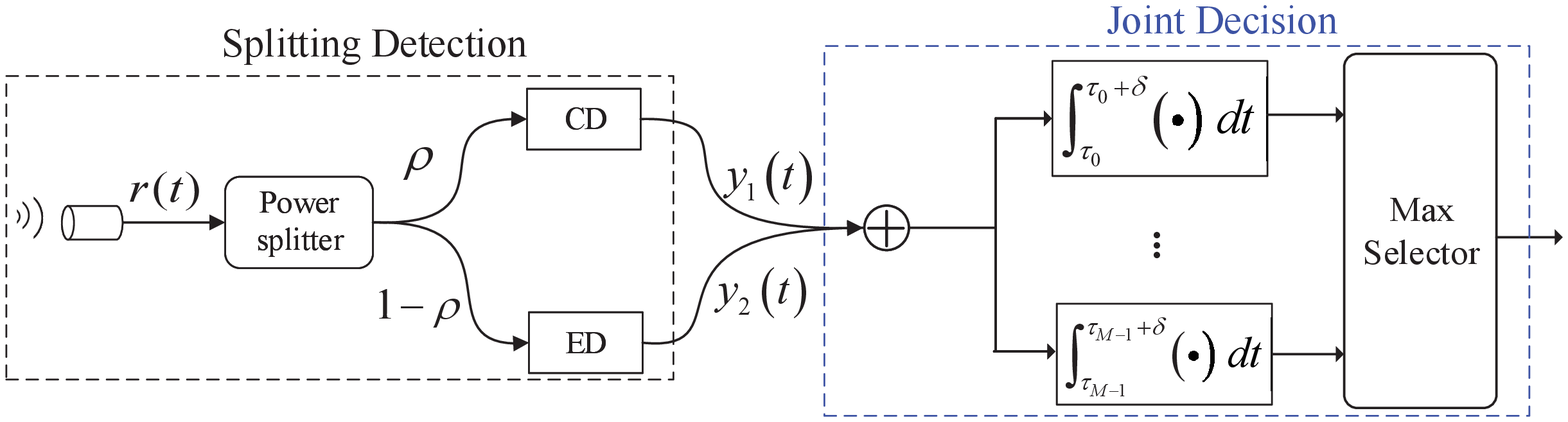}}\\
	\caption{Structures of the (a) SDSD receiver and (b) SDJD receiver.}
	\label{SDSD-JD}
\end{figure}

\subsubsection{SDSD}
The structure of SDSD is shown in Fig. \ref{SDSD}. The split signals are detected by CD and ED, and then are separately processed in two decision circuits to generate decision variables of ${\bf Y}_1$ and ${\bf Y}_2$. Based on the estimated channel coefficient, the received signal can be demodulated according to the ML-MMSE algorithm \cite{SDSD2017}.

\subsubsection{SDJD}
As shown in Fig. \ref{SDJD}, the received signal in the SDJD receiver is also split to CD and ED. Different from SDSD that generates two independent decision vectors, i.e., ${\bf Y}_1$ and ${\bf Y}_2$, the CD and ED signals in SDJD are summed. The combined signal is then processed in only one circuit to generate decision variables. With a max selector, the received signal can be readily demodulated. 
\newpage
\subsection{Signal and Channel Models}
\subsubsection{Transmitted signal}
 {Since pulse position modulation (PPM) is widely used in the pulse-based communication systems and can be received by both the CD and ED receivers \cite{b29,b30,b31}, we assume that the transmitter employs orthogonal $M$-ary PPM ($M$-PPM ) to transmit ultrasonic pulses. The transmitted monocycle waveform of one pulse is $p(t)$ with normalized energy $\int_{-\infty}^{\infty}\!p^2\!(t)dt\!=\!1$, and the pulse width is $T_p$. Let $T_c$ be the time interval of one chip for transmitting one pulse, the transmitted signal can be expressed~as
\begin{equation}
	s(t)=\sum_{j=0}^{\infty}\sqrt{E_p}p(t-j T_c-a_j \delta),
\end{equation}
where $j$ represents the index for the $j$th chip; $E_p$ represents the energy of one transmitted pulse; $\delta$ is the displacement of a pulse; $a_j$ denotes the index mapping from transmitted bits, e.g., for 4-PPM, the position indices of $\{0,1,2,3\}$ correspond to the transmitted bits of \{00, 01, 10, 11\}.}

\subsubsection{Channel model}
{The propagation of ultrasonic waves in tissues is affected by attenuation and small-scale fading. The attenuation is related to the amplitude attenuation coefficient $\alpha$ (in $\rm [np \cdot cm^{-1}]$) and the transmission distance $d$, and can be calculated by $P(d) = P_0 e^{-\alpha d}$, where $P_0$ represents the initial pressure \cite{Attenuation,uSDN}. The parameter $\alpha$ is a function of the central frequency of a channel as $\alpha=af^b$, where $a$ and $b$ are the attenuation parameters that characterize the tissue. Typical values of $\alpha$, $a$, and $b$ in different tissues can be found in \cite{e-health}.

According to the existing experimental results of intra-body modeling, the intra-body channel fading coefficient $h$ follows a generalized Nakagami distribution \cite{uSDN,Sonar,Experimental} with probability density function (PDF)
\begin{equation}
	f(h)=\frac{{2z{m^m}{h^{2zm - 1}}}}{{\Gamma (m){\Omega ^m}}}{e^{ - \frac{m}{\Omega }{h^{2z}}}},
	\label{eqNaka}
\end{equation}
where $m$, $\Omega$ and $z$ are the shaping, spreading, and generalization parameters of the generalized Nakagami distribution function; $\Gamma(\cdot)$ denotes the gamma function} \footnote{When $z=1$, the generalized Nakagami distribution becomes the Nakagami-$m$ distribution.}. 

It is worth noting that the multipath effect in the human body may cause the energy leak in one chip to adjacent chips or frames, i.e., inter-chip interference (ICI) or inter-frame interference (IFI). By properly setting the pulse duration and the chip duration, we can effectively mitigate the multipath effect, thereby avoiding ICI and IFI \cite{DS-USWB,UsIM1}. For simplicity, it is assumed that the ICI and IFI have been eliminated.
\vspace{0.3cm}
\subsubsection{Signal splitting}
{\color{black}The received signal can be expressed as
\begin{equation}
\vspace{-0.3cm}
	r(t)=h(t)s(t)+n_0(t),
	\label{eqRx}
\end{equation}
where $h(t)$ is the channel coefficient, and $n_0(t)$ is the additive white Gaussian noise having the first-order distribution of $\mathcal{N}\left(0,N_0/2\right)$. The SNR at the receiver is defined as $\gamma=h^2E_p/N_0$.

The received signal in the CD branch is multiplied by template signals $m_i(t)$, as shown in Fig.~\ref{CD}, where $i=0,\dots,M-1$. The template signals are generated according to the true transmitted pulse shape (known at the receiver) given by
\begin{equation}
\vspace{-0.3cm}
	m_i(t)=\sum_{j=0}^{\infty}\sqrt{E_p}p(t-j T_c-i \delta).
	\label{eq4}
\end{equation}
Then, the CD signal $y_{1i}(t)$ can be expressed as
\begin{equation}
\vspace{-0.3cm}
	y_{1i}(t)=\sqrt{\rho} r(t)m_i(t).
	\label{eqCD}
\end{equation}

The ED branch processes the received signal by a square-law device, as shown in Fig.~\ref{ED}, which collects the energy of the received signal and generates the ED signal $y_{2}(t)$ given by
\begin{equation}
	y_{2}(t) = \left(\sqrt{1-\rho}r(t)\right)^2.
	\label{eqED}
	\vspace{-0.3cm}
\end{equation}}
\vspace{-0.3cm}
\subsubsection{SDSD}
The signal processing of SDSD is illustrated in Fig. \ref{SDSD}. The CD and ED signals, i.e., $y_{1i}(t)$ and $y_2(t)$, are integrated at specific time intervals, separately. For the CD signal, when the $i$th template signal matches with the transmitted signal within the $j$th chip, i.e., $i=a_j$, the $i$th integral output is given by

\begin{equation}\nonumber
\begin{aligned}
\hspace{-6cm}y_{1i}=\int_{\tau_i}^{\tau_i+\delta}  \sqrt{\rho}r(t)m_i(t)dt
\end{aligned}
\end{equation}
\newpage
\begin{equation}
\begin{aligned}
&=\int_{\tau_i}^{\tau_i+\delta}  \sqrt{\rho}h(t)\sqrt{E_p}p(t-j T_c-a_j \delta)\sqrt{E_p}p(t-j T_c-i \delta)dt\\
&\qquad+\int_{\tau_i}^{\tau_i+\delta}  \sqrt{\rho}n_0(t)\sqrt{E_p}p(t-j T_c-i \delta)dt\\
&=h\sqrt{\rho} E_p+\sqrt{\rho E_p}n_{1i},
\end{aligned}
\end{equation}
where $\tau_i$ represents the start time of the $i$th $\delta$ interval within one chip.

When the $i$th template signal does not match with the transmitted signal, i.e., $i\neq a_j$, the template signal is orthogonal to the transmitted signal in this case, and thus we have
\begin{equation}
\begin{aligned}
y_{1i}=\sqrt{\rho E_p}n_{1i}.
\end{aligned}
\end{equation}

{\color{black}Therefore, the $i$th integral output in the CD branch is given by
\begin{equation}
	\begin{aligned}
		y_{1i}&=\int_{\tau_i}^{\tau_i+\delta}  \sqrt{\rho}r(t)m_i(t)dt=\left\{\begin{matrix}
			h\sqrt{\rho} E_p+\sqrt{\rho E_p}n_{1i} \quad i=a_j,\\
			\sqrt{\rho E_p}n_{1i} \qquad \quad \ \ \ \ \ \ i\neq a_j,
		\end{matrix}\right.
	\end{aligned}
	\label{eqSD-CD}
	\vspace{-0.3cm}
\end{equation}

For the ED signals, the $i$th integral is expressed as
\begin{equation}
	\vspace{-0.3cm}
	\begin{aligned}
		y_{2i}=\int_{\tau_i}^{\tau_i+\delta}r^2(t)dt=\left\{\begin{matrix}(1-\rho)\left({h^2}{E_p}{\rm{ + 2}}h\sqrt {{E_p}} {n_{2i}} + n_{2i}^2\right)\  \ i=a_j,\\
			(1-\rho) n_{2i}^2 \qquad \qquad \qquad \qquad\ \ \ \ \ \ \ i\neq a_j,
		\end{matrix}\right.
		\label{eqSD-ED}
	\end{aligned}
\end{equation}}
where $n_{1i}$ and $n_{2i}$ are independent identically distributed (i.i.d.) Gaussian random variables (RVs) according to $\mathcal{N}\left(0,\frac{N_0}{2}\right)$. It is shown that the square-noise $n^2_{2i}=\int_{\tau_i}^{\tau_i+\delta}w^2(t)dt$ can be decomposed into a sum of approximately $2\delta W_{rx}$ independent RVs, where $W_{rx}$ is the noise bandwidth that is equal to the bandwidth of the transmitted signal \cite{NoncoherentUWB}. For clarity, let $\delta W_{rx}=c$. The statistical model for $n^2_{2i}$ can be described as a central chi-squared PDF \cite{NoncoherentUWB}. When $c>40$, the central limit theorem applies for $n^2_{2i}$ with a Gaussian approximation \cite{NoncoherentUWB,LDR}. Then, $n^2_{2i}$ has a mean of $cN_0$ and a variance~of~$cN_0^2$. 

{\color{black}Let the transmitted signal for one pulse be represented by a vector ${\mathbf x}_i$, which has ``1'' at the $i$th position and zeros at other positions within one chip, i.e., ${\mathbf x}_i =[0,\dots, 1,\dots,0]$, where $i\in\{0,\dots,M-1\}$. As shown in Fig. \ref{SDSD}, the decision matrix in the SDSD receiver can be expressed as
\begin{equation}
	\bf{Y} = {\bf{SX}} + \left[ {\begin{array}{*{20}{c}}
			{\sqrt {\rho {E_p}} {n_{10}}}&{\cdots}&{\sqrt {\rho {E_p}} {n_{1i}}}&{\cdots}&{\sqrt {\rho {E_p}} {n_{1(M-1)}}}\\
			{\left( {1 - \rho } \right)n_{20}^2}&{\cdots}&\left( {1 - \rho } \right)\left(n_{2i}^2 + 2h\sqrt {{E_p}} {n_{2i}}\right)&{\cdots}&{\left( {1 - \rho } \right)n_{2(M-1)}^2}
	\end{array}} \right],
	\label{eqy1y2}
\end{equation}
where ${\bf{Y}}=[{{\bf Y}_1}\  {{\bf Y}_2}]^T$ is a $2\times M$ matrix, and ${\bf{S}}=\left[{h\sqrt \rho  {E_p}}\ {\left( {1 - \rho } \right){h^2}{E_p}}\right]^T$.}

The decision criterion based on ML-MMSE follows~\cite{SDSD2017}
\begin{equation}
	\langle {\hat i} \rangle  = \mathop {\arg \min }\limits_i
	{\left\| {{\bf{Y}} - {\bar{\bf{S}}}{\bf X}_i} \right\|^2},
	\label{eqSDd}
\end{equation}
where $\bar{\bf{S}}$ represents the estimated channel coefficients, specifically,
\begin{equation}
	\vspace{-0.3cm}
	\bar{\bf{S}}=\left[{{\bar h}\sqrt \rho  {E_p}}\ {\left( {1 - \rho } \right){{\bar h}^2}{E_p}}\right]^T,\vspace{-0.3cm}
\end{equation}
and where 
\begin{equation}
	{\bar h}=h+h_e,\vspace{-0.3cm}
	\label{eqChannel_error}
\end{equation}
and $h_e \sim \mathcal{N}(0,\sigma_e^2)$ is the channel estimation error, which is independent of the channel coefficient~\cite{Channel error}.

Then, the transmitted signal can be recovered from $\langle {\hat i} \rangle$ by the SDSD~receiver. Since the received signal $\bf{Y}$ exploits the available DoFs by using both the CD and ED receivers, SDSD can achieve a lower BER than either the CD or ED receiver.

\subsubsection{SDJD}
{\color{black}The integral outputs of the combined CD and ED signals, as shown in Fig. \ref{SDJD}, are given by
\begin{equation}
\begin{aligned}
I_i=\int_{\tau_i}^{\tau_i+\delta}\left(y_{1i}(t)+y_{2i}(t)\right)dt,
\end{aligned}
\label{eqJD}
\end{equation}
where $i=0,\dots,M-1$. According to \eqref{eqSD-CD} and \eqref{eqSD-ED}, $I_i$ is written as
{\setlength\abovedisplayskip{0.1cm}
\setlength\belowdisplayskip{0.1cm}
\begin{equation}
\begin{aligned}
{I_i} = \left\{ {\begin{array}{*{20}{l}}
{h\sqrt \rho  {E_p} + \sqrt {\rho {E_p}} {n_{1i}}{\rm{ + }}\left( {1 - \rho } \right){{\left({h^2}{E_p}{\rm{ + 2}}h\sqrt {{E_p}} {n_{2i}} + n_{2i}^2 \right)}}\ \ i = {a_j},}\\
{\sqrt {\rho {E_p}} {n_{1i}} + \left( {1 - \rho } \right)n_{2i}^2\qquad  \qquad\qquad \qquad \qquad \qquad \quad\ \ \ i \ne {a_j}.}
\end{array}} \right.
\label{eqIi}
\end{aligned}
\end{equation}}
The received signal can be estimated by a maximum selector, based on the ML criterion,~i.e.,
\begin{equation}
\langle\hat{i}\rangle= \mathop {\arg \max}\limits_i\ \{I_{i}\}.
\label{eqJDd}
\end{equation}
The joint decision in \eqref{eqIi} implements the receive diversity by combining the detected signals in different DoFs of CD and ED, and thus improves BER performance.}

{\color{black}\section{Performance Analysis}} \label{sect: performance}	
In this section, we analyze the signal space and complexity of different receivers, and we derive the theoretical BER and channel capacity expressions for the SDSD and SDJD receivers using different modulation schemes, including $M$-PPM and PPM with spreading codes.

\begin{figure}[!t]
	\centering{\includegraphics[width=2in]{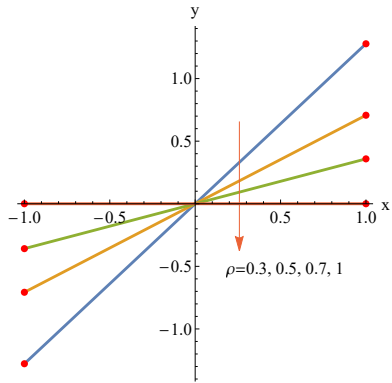}}
	\caption{Received signal space in splitting detection with different values of $\rho$, for 2-PPM.}
	\label{Space}
\end{figure}

\begin{table}[!t]
	\color{black}
	\caption{Parameters of different detection in signal space}
	\vspace{-0.3cm}
	\setlength{\tabcolsep}{3pt}
	\centering
	\small
	\begin{tabular} {c|c|c|c}
		\hline\hline
		Detection & Splitting ratio $\rho$ &DoFs & Distance \\
		\hline\hline
		CD &  1 & 1 ($x$ axis) & 2 \\
		\hline
		ED &  0 & 1 ($y$ axis) & 2  \\
		\hline
		Splitting Detection (SDSD and SDJD) &  $0<\rho<1$ &  2 & $2\sqrt{1+\frac{(1-\rho)^2}{\rho}}$ \\
		\hline
	\end{tabular}
	\vspace{-0.3cm}
	\label{tab1}
\end{table}

\subsection{Signal Space} 
{\color{black}Assuming that the received signal is transmitted over a noiseless channel, and the received pulses have normalized energy. For $M$-PPM with splitting detection, the signal can be positioned in a two-dimensional plane. The received signal space is determined by the \emph{pulse position} and \emph{energy position}, which are referred to as $x$ and $y$ in Fig. \ref{Space}, respectively. Let $x_i=2i-1$ for $i=1,2,\dots,M/2$, and $x_i=-x_{i-M/2}$ for $i=M/2+1,\dots,M$. Let $y_i=k x_i$, where $k=(1-\rho)/{\sqrt{\rho}}$. Take 2-PPM as an example.} Fig. \ref{Space} depicts the received \emph{signal space} in splitting detection that is shown with the orthogonal axes by $x$ and $y$, with different values of $\rho$. The distance between two signal positions is calculated by $d=\sqrt{(x_2-x_1)^2+(y_2-y_1)^2}$. The parameters of different detection in signal space are given in Tab.~\ref{tab1}.

\emph{Remark 1:} From Tab. \ref{tab1}, we can find that the distance between two symbols in the splitting receivers is larger than those in the CD and ED receivers. An increase in DoFs further increases the Euclidean distance of the symbols in both the SDSD and SDJD receivers, which is beneficial to improve BER~performance.

\subsection{Receiver Complexity}
Figure \ref{SDSD-JD} compares the structures of SDSD and SDJD. The SDSD receiver requires the CD and ED receivers to generate separate ${\bf Y}_1$ and ${\bf Y}_2$, while SDJD only needs CD and ED to implement diversity combining, which greatly reduces the hardware complexity. Additionally, from the CD receiver in Fig. \ref{CD}, we can find that the SDJD receiver can be easily implemented by adding just a few additional simple operations to the CD receiver, including square-law and summation operations. Hence the structure of the SDJD receiver is slightly more complex than that of the CD and ED receivers. However, the SDJD receiver has a relatively lower BER than both the CD and ED receivers, which will be proven in Sections \ref{Sect: BER} and \ref{Sect:Simulation}.

Suppose the computational complexity is calculated by multiplications. On the one hand, it can be inferred from \eqref{eqSDd} that the complexity of the decision for SDSD is $\mathcal{O}(M)$ for $M$-PPM. While there is no multiplication for the SDJD receiver as shown in \eqref{eqJDd}, leading to much lower complexity than SDSD. On the other hand, the SDSD receiver requires exact channel estimation, which consumes additional energy and operations. In contrast, owing to the characteristics of ultrasonic PPM, it is not necessary to perform channel estimation for the SDJD receiver. The complexity of SDJD in terms of multiplication and hardware is lower than that of SDSD, which makes the SDJD receiver easier to implement.

{\color{black}Therefore, the SDJD receiver is more suitable for implant devices than the CD or ED receiver and the SDSD receiver.}

\subsection{Channel Capacity}
{\color{black}Assuming that the channel capacity with input signals is restricted to an equiprobable $M$-PPM constellation, i.e., $p({\bf X}={\bf x})=1/M$, where ${\bf x}\in\{{\bf x}_0,\dots,{\bf x}_{M-1}\}$.}
\subsubsection{Capacity of SDSD}
For the SDSD receiver, the detected signal vectors for one received pulse in the CD and ED receivers are expressed as ${\bf Y}_1=[y_{10},\dots,y_{1(M-1)}]$ and ${\bf Y}_2=[y_{20},\dots,$ $y_{2(M-1)}]$, respectively, corresponding to $M$ positions within one chip. Let $[{y}_{1i}\ {y}_{2i}]^T$ represent the positions that contain ultrasonic pulse. The non-pulse positions are given by $[{y}_{1j}\ {y}_{2j}]^T$, where $j=0,\dots,M-1$, and $j\neq i$. The distributions of ${y}_{1i},{y}_{2i},{y}_{1j},{y}_{2j}$ are respectively given by
\begin{equation}
{y}_{1i} \sim \mathcal{N}\left(\mu_{11},\sigma _{11}^2 \right),\quad 
{y}_{1j} \sim \mathcal{N}\left(\mu_{12},\sigma _{12}^2 \right),\quad 
{y}_{2i} \sim \mathcal{N}\left(\mu_{21},\sigma _{21}^2 \right),\quad 
{y}_{2j} \sim \mathcal{N}\left(\mu_{22},\sigma _{22}^2 \right),
\label{eqY_i,j}
\end{equation}
where the mean and variance values for these Gaussian distributions are given in Tab. \ref{tab2}.

\begin{table}[!t]
	\color{black}
	\caption{Statistical parameters of the received signal in the SDSD receiver.}
	\vspace{-0.3cm}
	\setlength{\tabcolsep}{3pt}
	\centering
	\small
	\begin{tabular} { c|c||c|c}
		\hline\hline
		Mean & Value & Variance & Value\\
		\hline\hline
		$\mu _{11}$ &  $h\sqrt \rho  {E_p}$ & $\sigma _{11}^2$ & $\frac{{\rho {E_p}{N_0}}}{2}$\\
		\hline
		$\mu _{12}$ &  $0$ & $\sigma _{12}^2$ & $\frac{{\rho {E_p}{N_0}}}{2}$ \\
		\hline
		$\mu _{21}$ &  $\left( {1 - \rho } \right){h^2}{E_p}+\left( {1 - \rho } \right)c{N_0}$ & $\sigma _{21}^2$ & $2{{\left( {1 - \rho } \right)}^2}{h^2}{E_p}{N_0} + {{\left( {1 - \rho } \right)}^2}c{N_0}^2$ \\
		\hline
		$\mu _{22}$ &  $\left( {1 - \rho } \right)c{N_0}$ & $\sigma _{22}^2$ & ${{\left( {1 - \rho } \right)}^2}c{N_0^2}$ \\
		\hline
	\end{tabular}
	\vspace{-0.3cm}
	\label{tab2}
\end{table}

Due to symmetry of $M$-PPM, according to the definition of channel capacity \cite{Info.Theo.}, the capacity (in [bits/channel use]) of $M$-PPM can be calculated as \cite{CapacityPPM} 
\begin{equation}
\begin{aligned}
C_{\rm {\small SDSD}} =\log _{2} M-E_{{\bf Y} | {\bf x}_{1}} \left\{\log _{2}\left[\frac{\sum_{i=1}^{M} p\left({\bf Y} | {\bf x}_{i}\right)}{p\left({\bf Y} | {\bf x}_{1}\right)}\right]\right\}.
\end{aligned}
\label{eqCapacity}
\end{equation}

\begin{table}[!t]
	\color{black}
	\caption{Statistical parameters of the received signal in the SDJD receiver.}
	\vspace{-0.3cm}
	\small
	\centering
	\begin{tabular} { c|c||c|c}
		\hline\hline
		Mean & Value & Variance & Value\\
		\hline\hline
		$\mu _{1}$ &  $h\sqrt \rho  {E_p}+{h^2}(1 - \rho ){E_p} +\left( {1 - \rho } \right)c{N_0}$ & $\sigma _{1}^2$ & $\frac{{\rho {E_p}{N_0}}}{2} + 2{h^2}{{(1 - \rho )}^2}{E_p}{N_0} + {{{(1 - \rho )}^2}N_0^2c}$\\
		\hline
		$\mu _{2}$ &  $\left( {1 - \rho } \right)c{N_0}$ & $\sigma _{2}^2$ & $\frac{{\rho {E_p}{N_0}}}{2} + {{{(1 - \rho )}^2}N_0^2c}$ \\
		\hline
	\end{tabular}
	\vspace{-0.6cm}
	\label{tab3}
\end{table}
Based on \eqref{eqY_i,j}, the conditional PDF $p({\mathbf Y}|{\mathbf x}_i)$ is given by
\newpage
\begin{equation}
\begin{aligned}
p({\bf{Y}}|{{\bf{x}}_i}) = &\frac{1}{{{{\left( {2\pi }\sigma _{11} \right)}^M} }}\exp \left[ { - \frac{{{{\left( {{y_{1i}} - {\mu _{11}}} \right)}^2}}}{{2\sigma _{11}^2}}} \right]\mathop \Pi \limits_{j= 1\hfill\atop
	j\ne i\hfill}^M \exp \left[ { - \frac{{{{ {{y_{1j}}}}^2}}}{{2\sigma _{11}^2}}} \right]\\
&\times\frac{1}{\sqrt {\sigma _{21}^2\sigma _{22}^{2(M - 1)}}} \exp \left[ { - \frac{{{{\left( {{y_{2i}} - {\mu _{21}}} \right)}^2}}}{{2\sigma _{21}^2}}} \right]\mathop \Pi \limits_{j= 1\hfill\atop
	j\ne i\hfill}^M \exp \left[ { - \frac{{{{\left( {{y_{2j}} - {\mu _{22}}} \right)}^2}}}{{2\sigma _{22}^2}}} \right].
\end{aligned}
\label{equy1}
\end{equation}
Then, we obtain
\begin{equation}
\begin{aligned}
\frac{{p({\bf{Y}}|{{\bf{x}}_i})}}{{p({\bf{Y}}|{{\bf{x}}_1})}} &= \exp \left[ \begin{array}{l}
\frac{{y_{1i}^2}}{{2\sigma _{11}^2}} - \frac{{{{\left( {{y_{1i}} - {\mu _{11}}} \right)}^2}}}{{2\sigma _{11}^2}} - \frac{{y_{11}^2}}{{2\sigma _{11}^2}} + \frac{{{{\left( {{y_{11}} - {\mu _{11}}} \right)}^2}}}{{2\sigma _{11}^2}}\\
+ \frac{{{{\left( {{y_{2i}} - {\mu _{22}}} \right)}^2}}}{{2\sigma _{22}^2}} - \frac{{{{\left( {{y_{2i}} - {\mu _{21}}} \right)}^2}}}{{2\sigma _{21}^2}} - \frac{{{{\left( {{y_{21}} - {\mu _{22}}} \right)}^2}}}{{2\sigma _{22}^2}} + \frac{{{{\left( {{y_{21}} - {\mu _{21}}} \right)}^2}}}{{2\sigma _{21}^2}}
\end{array} \right]\\
&=\exp \left[ {\sqrt {{m_0}} \left( {{v_{1i}} - {v_{11}}} \right) + \frac{1}{2}\left( {u_{2i}^2 - v_{2i}^2 - u_{21}^2 + v_{21}^2} \right)} \right],
\end{aligned}
\label{eqYratio}
\end{equation}
where ${v_{1i}} = \left({y_{1i}} - {\mu _{11}}\right)/\sigma _{11}$, ${u_{1i}} = \left({y_{2i}} - {\mu _{22}}\right)/\sigma _{22}$, ${v_{2i}} = \left({y_{2i}} - {\mu _{21}}\right)/\sigma _{21}$, and $m_0=2\gamma$. Given ${\bf x}_1$, the distributions are given by
\begin{equation}
v_{11}\sim\mathcal{N}(\sqrt{m_0},1),\quad
v_{1i}\sim\mathcal{N}(0,1),\quad
v_{21}\sim\mathcal{N}\left(\sqrt{m_{20}},\frac{m_{20}+m_{21}}{m_{21}}\right),\quad
v_{2i}\sim\mathcal{N}(0,1),
\label{value}
\end{equation}
and the distribution of $u_{2i}$ can be represented by $u_{2i}=\sqrt{m_{21}/\left(m_{20}+m_{21}\right)}(v_{2i}-\sqrt{m_{20}})$. Specifically,
\begin{equation}
{m_{20}} = \frac{{{{\left( {{\mu _{21}} - {\mu _{22}}} \right)}^2}}}{{\sigma _{22}^2}} = \frac{{{\gamma ^2}}}{c},\qquad
{m_{21}} = \frac{{{{\left( {{\mu _{21}} - {\mu _{22}}} \right)}^2}}}{{\sigma _{21}^2 - \sigma _{22}^2}} = \frac{\gamma }{2}.
\end{equation}
Substituting \eqref{eqYratio} and \eqref{value} into \eqref{eqCapacity}, we can express the channel capacity as
\begin{equation}
C_{\rm {\small SDSD}} = {\log _2}M\! -\! {E_{{\bf{v}}|{{\bf{x}}_1}}}\left\{{\log _2}\sum\limits_{i = 1}^M {\exp \left[\!\! \begin{array}{c}
	\sqrt {{m_0}} \left( {{v_{1i}} - {v_{11}}} \right) + \frac{{{m_{21}}}}{{{m_{20}} + {m_{21}}}}\sqrt {{m_{20}}} \left( {{v_{2i}} - {v_{21}}} \right)\\
	\qquad\ + \frac{{{m_{20}}}}{{{m_{20}} + {m_{21}}}}\frac{{v_{2i}^2 - v_{21}^2}}{2}
	\end{array} \!\!\right]}\right\}.
\end{equation}

\emph{Remark 2:} The channel capacities of the CD and ED receivers can be shown to be
\begin{equation}
{C_{{\rm{CD}}}} = {\log _2}M - {E_{{\bf{v}}|{{\bf{x}}_1}}}\left\{{\log _2}\sum\limits_{i = 1}^M {\exp \left[ {\sqrt {{m_0}} \left( {{v_{1i}} - {v_{11}}} \right)} \right]}\right\},
\end{equation}
\begin{equation}
\vspace{-0.3cm}
{C_{{\rm{ED}}}}\!\! =\!\! {\log _2}M \!\!-\!\! {E_{{\bf{v}}|{{\bf{x}}_1}}}\left\{{\log _2}\sum\limits_{i = 1}^M {\exp \left[ {\frac{{{m_{21}}}}{{{m_{20}} + {m_{21}}}}\sqrt {{m_{20}}} \left( {{v_{2i}} \!\!-\!\! {v_{21}}} \right) + \frac{{{m_{20}}}}{{{m_{20}} \!\!+\!\! {m_{21}}}}\frac{{v_{2i}^2 \!\!-\!\! v_{21}^2}}{2}} \right]}\right\}.
\end{equation}
The channel capacities of the CD and ED receivers are determined by SNR. Since the SNRs in the independent CD and ED receivers of SDSD are independent of the splitting ratio, the channel capacity of the SDSD receiver  does not depend on the splitting ratio.

\subsubsection{Capacity of SDJD}
Let the detected signal vector in the SDJD receiver be expressed as ${\bf Z}=[z_0,\dots,z_{M-1}]$, which corresponds to $M$ positions within one chip. Let $z_i$ represent the position that contains the ultrasonic pulse. The non-pulse positions are given by $z_j$, where $j=0,\dots,M-1$, and $j\neq i$. The distributions of $z_i$ and $z_j$ are respectively given by
\begin{equation}
\vspace{-0.3cm}
\begin{matrix}
z_i \sim \mathcal{N}\left(\mu_1,\sigma _1^2 \right),\qquad z_j \sim \mathcal{N}\left(\mu_2,\sigma _2^2 \right)\
\end{matrix},
\label{eqT_i,j}
\end{equation}
where the mean and variance values are shown in Tab. \ref{tab3}. The conditional PDF $p({\bf Z}|{\bf x}_i)$ is given~by
\begin{equation}
\begin{aligned}
p({\bf{Z}}|{{\bf{x}}_i}) = \frac{1}{{\sqrt {{{\left( {2\pi } \right)}^M}\sigma _1^2\sigma _2^{2(M - 1)}} }}\exp \left[ { - \frac{{{{\left( {{z_i} - {\mu _1}} \right)}^2}}}{{2\sigma _1^2}}} \right]\mathop \Pi \limits_{ j= 1\hfill\atop
	j\ne i\hfill}^M \exp \left[ { - \frac{{{{\left( {{z_j} - {\mu _2}} \right)}^2}}}{{2\sigma _2^2}}} \right].
\end{aligned}
\label{eq24}
\end{equation}
Then, we obtain
\begin{equation}
\vspace{-0.3cm}
\frac{{p({\bf Z}|{{\bf x}_i})}}{{p({\bf Z}|{{\bf x}_1})}} =\exp \left[ {\frac{1}{2}\left( {v_i^2 - u_i^2 + u_1^2 - v_1^2} \right)} \right],
\label{eqIratio}
\end{equation}
where ${v_1}\sim{\cal N}\left( {\frac{{{\mu _1} - {\mu _2}}}{{{\sigma _2}}},\frac{{\sigma _1^2}}{{\sigma _2^2}}} \right)$ and ${v_i}\sim{\cal N}\left( {0,1} \right)$. Let
{\setlength\abovedisplayskip{0cm}
\setlength\belowdisplayskip{0cm}
\begin{equation}
\begin{aligned}
{m_0} = \frac{{{{\left( {{\mu _1} - {\mu _2}} \right)}^2}}}{{\sigma _2^2}}=\frac{{{{\left( {\sqrt {\frac{\rho }{{{{\left( {1 - \rho } \right)}^2}{h^2}}}}  + 1} \right)}^2}}}{{\frac{\rho }{{2\gamma {{\left( {1 - \rho } \right)}^2}{h^2}}} + \frac{c}{{{\gamma ^2}}}}},
\end{aligned}
\label{eqm0}
\end{equation}}
{\setlength\abovedisplayskip{0cm}
	\setlength\belowdisplayskip{0cm}
	\begin{equation}
\begin{aligned}
{m_1} = \frac{{{{\left( {{\mu _1} - {\mu _2}} \right)}^2}}}{{\sigma _1^2 - \sigma _2^2}}=\frac{{{{\left( {\sqrt {\frac{\rho }{{{{\left( {1 - \rho } \right)}^2}{h^2}}}}  + 1} \right)}^2}\gamma }}{2},
\end{aligned}
\label{eqm1}
\end{equation}}
we have ${v_1}\sim{\cal N}\left( {\sqrt {{m_0}} ,\frac{{{m_0} + {m_1}}}{{{m_1}}}} \right)$. Using \eqref{eqCapacity} and \eqref{eqIratio}, we write the channel capacity as
\newpage
\begin{equation}
\vspace{-0.3cm}
\begin{aligned}
C_{\rm \small SDJD} &\!\!=\!\! {E_h}\left\{ {{\log }_2}M\!\! -\!\! {E_{{\bf{Z}}|{{\bf{x}}_1}}}\left\{\!\!{{\log }_2}\sum\limits_{i = 1}^M {\exp \left[\!\! {\frac{{{{\left( {{z_1} \!\!-\!\! {\mu _1}} \right)}^2} \!\!-\!\! {{\left( {{z_i} \!\!-\!\! {\mu _1}} \right)}^2}}}{{2\sigma _1^2}} + \frac{{{{\left( {{z_i} \!\!-\!\! {\mu _2}} \right)}^2} \!\!-\!\! {{\left( {{z_1} \!\!- \!\!{\mu _2}} \right)}^2}}}{{2\sigma _2^2}}}\! \right]} \! \right\}\!\right\}\\
&\!\!=\!\!{E_h}\left\{{\log _2}M \!\!-\!\! {E_{{\bf{v}}|{{\bf{x}}_1}}}\left\{\!\!{\log _2}\sum\limits_{i = 1}^M {\exp \left[ {\frac{{{m_1}}}{{{m_0} + {m_1}}}\sqrt {{m_0}} \left( {{v_i} \!\!-\!\! {v_1}} \right) + \frac{{{m_0}}}{{{m_0} + {m_1}}}\frac{{v_i^2 \!\!-\!\! v_1^2}}{2}} \right]}\!\right\}\!\right\}.
\end{aligned}
\label{eqCapacityJD}
\end{equation}
The $M$-dimension expectation in the right-side of \eqref{eqCapacityJD} is complicated. We will calculate it via Monte Carlo simulations. 

\emph{Remark 3:} Since the CD and ED signals are summed in the SDJD receiver, the channel capacity is characterized by $m_0$ and $m_1$, which are given in \eqref{eqm0} and \eqref{eqm1}, respectively. We can observe that both  $m_0$ and $m_1$ contain  $\rho/((1-\rho)^2h^2)$. Therefore, the channel capacity of SDJD is affected by the splitting ratio, which is different from SDSD (see \emph{Remark~2}).
\vspace{-0.6cm}
\subsection{BER} \label{Sect: BER}
\vspace{-0.3cm}
\subsubsection{BER for SDSD}
For $M$-PPM, let the transmitted signal be in the first interval within one chip in the SDSD receiver. According to \eqref{eqSDd}, the correct decision occurs when $ {{\left\| {{\bf{Y}} - \bar{\bf{S}}{{\bf{X}}_0}} \right\|}^2} < {{\left\| {{\bf{Y}} - \bar{\bf{S}}{{\bf{X}}_i}} \right\|}^2} ,i=1,\dots,M-1$. Hence, the probability of a correct decision is calculated as
\begin{equation}
\vspace{-0.3cm}
P_c=P\left[ {{{\left\| {{\bf{Y}} - \bar{\bf{S}}{{\bf{X}}_1}} \right\|}^2} > {{\left\| {{\bf{Y}} - \bar{\bf{S}}{{\bf{X}}_0}} \right\|}^2},\dots,{{\left\| {{\bf{Y}} - \bar{\bf{S}}{{\bf{X}}_{M-1}}} \right\|}^2} > {{\left\| {{\bf{Y}} - \bar{\bf{S}}{{\bf{X}}_0}} \right\|}^2}|{{\bf{x}}_0}\ {\rm{sent}}} \right].
\end{equation}
Based on \eqref{eqy1y2}, we can find that all events $\left\{{{\left\| {{\bf{Y}} - \bar{\bf{S}}{{\bf{X}}_i}} \right\|}^2} > {{\left\| {{\bf{Y}} - \bar{\bf{S}}{{\bf{X}}_0}} \right\|}^2}\right\}$ contain the variable $2\rho \bar hE_p^{\frac{3}{2}}{n_{10}} + 4{{\left( {1 - \rho } \right)}^2}{{\bar h}^2}hE_p^{\frac{3}{2}}{n_{20}} + 2{{\left( {1 - \rho } \right)}^2}{{\bar h}^2}{E_p}n_{20}^2$, which is denoted by $n_1$; thus, they are not independent. To make these events independent, we can condition on $n_1$. Then, we have
\begin{equation}
\vspace{-0.3cm}
\begin{aligned}
{P_c} = \int_{ - \infty }^\infty\left(P\left( \begin{array}{l}
	2h\bar h\rho E_p^2 + 2{\left( {1 - \rho } \right)^2}{{\bar h}^2}{h^2}E_p^2 + n> \\
	  \quad 2\rho \bar hE_p^{\frac{3}{2}}{n_{11}}+ 2{\left( {1 - \rho } \right)^2}{{\bar h}^2}{E_p}n_{21}^2|{n_1} = n,{{\bf{x}}_0}\;{\rm{sent}}
	\end{array} \right)\right)^{M-1}{p_{{n_1}}}\left( n \right) dn.
\end{aligned}
\end{equation}
Recalling the analysis in Section \ref{sect: Received Signal}, since $n_{10},n_{11},n_{20},n_{21}$ are i.i.d. RVs following $\mathcal{N}(0,\frac{N_0}{2})$ and $n_{20}^2,n_{21}^2$ are approximated as i.i.d. Gaussian RVs following $\mathcal{N}\left(cN_0,cN_0^2\right)$, we have
\begin{equation} \nonumber
\begin{aligned}
P_{c}=\int_{ - \infty }^\infty\left(1 - Q\left( {\frac{{2h\bar h\rho E_p^2 + 2{{\left( {1 - \rho } \right)}^2}{{\bar h}^2}{h^2}E_p^2 + n - 2{{\left( {1 - \rho } \right)}^2}{{\bar h}^2}{E_p}c{N_0}}}{{\sqrt {{{2\left( {\rho \bar hE_p^{\frac{3}{2}}} \right)}^2}{N_0} + {{\left( {2{{\left( {1 - \rho } \right)}^2}{{\bar h}^2}{E_p}} \right)}^2}c{N_0^2}} }}} \right)\right)^{M-1}\!\!\!\!\!{p_{{n_1}}}\left( n \right) dn
\end{aligned}
\end{equation}
\newpage
\begin{equation}
\begin{aligned}
= \int_{ - \infty }^\infty\left(1 - Q\left( {\frac{{2\bar h\rho /h + 2{{\left( {1 - \rho } \right)}^2}{{\bar h}^2} + n - 2{{\left( {1 - \rho } \right)}^2}{{\bar h}^2}c/\gamma }}{{\sqrt {2{\rho ^2}{{\bar h}^2}/{h^2}\gamma  + 4{{\left( {1 - \rho } \right)}^4}{{\bar h}^4}c/{\gamma ^2}} }}} \right)\right)^{M-1}{p_{{n_1}}}\left( n \right) dn,
\end{aligned}
\label{eqPc}
\end{equation}
where $p_{n_1}(n)$ is the PDF of $\mathcal{N}\left(\frac{2{{\left( {1 - \rho } \right)}^2}{{\bar h}^2}c}{\gamma},\frac{2{\rho ^2}{{\bar h}^2}}{{h^2}\gamma}  + \frac{8{{\left( {1 - \rho } \right)}^4}{{\bar h}^4}}{\gamma}  + \frac{4{{\left( {1 - \rho } \right)}^4}{{\bar h}^4}c}{\gamma ^2}\right)$. Note that although the closed-form $P_c$ is intractable, it can be calculated numerically.

Then, the probability of error detection is given by $P_e=1-P_c$. Assuming that ${\bf x}_0$ corresponds to a transmission bit sequence of length $k$ with a ``0'' at the first component. Due to the symmetry of the constellation for orthogonal $M$-PPM, when ${\bf x}_0$ is sent, the probabilities of receiving other signals of ${\bf x}_i,\ i=1,\dots,M-1$ are equal. Therefore, for any $1\leqslant i\leqslant M-1$,
\begin{equation}
\begin{aligned}
P\left[ {{{\bf{x}}_i}{\rm{\  received}}|{{\bf{x}}_0}{\rm{\  sent}}} \right] = \frac{{{P_e}}}{{M - 1}} = \frac{{{P_e}}}{{{2^k} - 1}}.
\label{eqSDSD-BER1}
\end{aligned}
\vspace{-0.3cm}
\end{equation}
The error probability at the first component of ${\bf x}_0$ is the probability of detecting an ${\bf x}_m$ corresponding a sequence with a ``1'' at the first component. Since there are $2^{k-1}$ such transmission sequences, the average BER is given by \cite{DigiCom}
\begin{equation}
\begin{aligned}
{P_b} = \frac{{{2^{k - 1}}}}{{{2^k} - 1}}{P_e}.
\label{eqSDSD-BER2}
\end{aligned}
\vspace{-0.3cm}
\end{equation}
\vspace{-0.9cm}
\subsubsection{BER for SDJD}
Also, let us consider $M$-PPM with the signal transmitting in the first interval. According to decision criterion as shown in \eqref{eqJDd}, the detector makes a correct decision if $I_{0} > {I_i}$, $i=1,\dots,M-1$. Therefore, a correct decision probability is given by
\begin{equation}
P_c={P}\left( { I_0 > {I_1},\dots,I_0 > {I_{M-1}}}\big|{{\bf{x}}_0}\ {\rm{sent }} \right).
\end{equation}
According to \eqref{eqIi}, it can be found that all events $\left\{I_{1} > {I_i}\right\}$ contain the variable $\sqrt {\rho {E_p}} {n_{10}} \!+\! 2\left( {1 - \rho } \right)h\sqrt {{E_p}} {n_{20}} + \left( {1 - \rho } \right)n_{20}^2$, which is again represented by $n_1$. We can condition on $n_1$ to make these events independent. Thus, we have
\begin{equation}\nonumber
\begin{aligned}
{P_c} = {\int_{ - \infty }^\infty  {\left( {P\left( \begin{array}{l}
			h\sqrt \rho  {E_p} + {h^2}\left( {1 - \rho } \right){E_p} + n > \vspace{-0.3cm}\\
			\quad\sqrt {\rho {E_p}} {n_{11}} + \left( {1 - \rho } \right)n_{21}^2|{n_1} = n,{{\bf{x}}_0}\;{\rm{sent}}
			\end{array} \right.} \right)} ^{M - 1}}{p_{{n_1}}}\left( n \right)dn
\end{aligned}
\end{equation}
\newpage
\begin{equation}
\begin{aligned}
= \int_{ - \infty }^\infty  {{{\left( {1 - Q\left( {\frac{{\sqrt \rho   + h\left( {1 - \rho } \right) + n - \frac{{\left( {1 - \rho } \right)ch}}{\gamma }}}{{\sqrt {\frac{\rho }{{2\gamma }} + \frac{{{{\left( {1 - \rho } \right)}^2}c{h^2}}}{{{\gamma ^2}}}} }}} \right)} \right)}^{M - 1}}} {p_{{n_1}}}\left( n \right)dn,
\end{aligned}
\end{equation}
where $p_{n_1}(n)$ is the PDF of $\mathcal{N}\left(\frac{\left( {1 - \rho } \right)ch}{\gamma},\frac{\rho}{2\gamma} + \frac{{{2\left( {1 - \rho } \right)}^2h^2}}{\gamma} + \frac{{{\left( {1 - \rho } \right)}^2}c{h^2}}{\gamma ^2}\right)$. Finally, adopting the analysis  similar to SDSD, we can calculate the average BER of the SDJD receiver by \eqref{eqSDSD-BER2}.

\subsubsection{Comparison between SDSD and SDJD}
To obtain more insights from comparing the BER between SDSD and SDJD, for simplicity, we take 2-PPM as an example. The BER conditioned on the channel coefficients $h$ and $\bar h$ of SDSD, is given by
\begin{equation}
\begin{aligned}
P_{\!esd}&\!=\!\frac{1}{2}P\!\left[\! {{{\left\| {{\bf{Y}}\! -\! {\bar{\bf{S}}}{{\bf{X}}_1}} \right\|}^2} \!\!<\! {{\left\| {{\bf{Y}}\! -\! {\bar{\bf{S}}}{{\bf{X}}_0}} \right\|}^2}|{{\bf{x}}_0}{\rm{\ sent}}} \right] \!+\! \frac{1}{2}P\left[ {{{\left\| {{\bf{Y}} \!-\! {\bar{\bf{S}}}{{\bf{X}}_0}} \right\|}^2}\! <\! {{\left\| {{\bf{Y}} \!-\! {\bar{\bf{S}}}{{\bf{X}}_1}} \right\|}^2}|{{\bf{x}}_1}{\rm{\ sent}}} \right]\!\\
&=Q\left( {\frac{{\left( {\sqrt {\frac{\rho }{{{{\left( {1 - \rho } \right)}^2}{{\bar h}^2}}}\frac{\rho }{{{{\left( {1 - \rho } \right)}^2}{h^2}}}}  + 1} \right)\gamma }}{{\sqrt {\frac{\rho }{{{{\left( {1 - \rho } \right)}^2}{{\bar h}^2}}}\frac{\rho }{{{{\left( {1 - \rho } \right)}^2}{h^2}}}\gamma  + 2\gamma  + 2c} }}} \right).
\end{aligned}
\label{eqSDSD}
\end{equation}
For SDJD, the BER conditioned on the channel coefficient $h$ is calculated as
\begin{equation}
\begin{aligned}
P_{ejd}&=\frac{1}{2}P\left[ {{I_1} > {I_0}|{{\bf{x}}_0}\;{\rm{sent}}} \right] + \frac{1}{2}P\left[ {{I_0} > {I_1}|{{\bf{x}}_1}\;{\rm{sent}}} \right]\\
&=Q\left( {\frac{{\left( {\sqrt \rho  {\rm{ + }}\left( {1 - \rho } \right)h} \right)\gamma }}{{\sqrt {\rho \gamma  + 2{{\left( {1 - \rho } \right)}^2}{h^2}\gamma  + 2{{\left( {1 - \rho } \right)}^2}{h^2}c} }}} \right)
\end{aligned}
\label{eqSDJD}
\end{equation}
\begin{equation}
\hspace{-1.5cm}=Q\left( {\frac{{\left( {\sqrt {\frac{\rho }{{{{\left( {1 - \rho } \right)}^2}{h^2}}}} {\rm{ + }}1} \right)\gamma }}{{\sqrt {\frac{\rho }{{{{\left( {1 - \rho } \right)}^2}{h^2}}}\gamma  + 2\gamma  + 2c} }}} \right).
\label{eqSDJD-2}
\end{equation}
The BER of the CD or ED receiver can be derived by substituting $\rho=1$ or $\rho=0$ into \eqref{eqSDJD},~i.e.,
\begin{equation}
\begin{aligned}
\begin{array}{l}
\left\{ \begin{array}{l}
{P_{CD}} = {Q\left( {\sqrt \gamma  } \right)},\\
{P_{ED}} = {Q\left( {\frac{\gamma }{{\sqrt {2\gamma  + 2c} }}} \right)}.
\end{array} \right.
\end{array}
\end{aligned}
\end{equation}

\emph{Remark 4}: Comparing \eqref{eqSDSD} and \eqref{eqSDJD-2}, we can find that the BER of SDSD is affected by the estimated and practical channel coefficients, i.e., $\bar h$ and $h$; however, the BER of SDJD is only affected by the practical channel coefficient. Therefore, in the case of imperfect channel estimation, the SDJD receiver is superior to the SDSD receiver.

{\color{black}\subsubsection{Optimal splitting ratio}}
Since the Gaussian $Q$-function is a monotonically decreasing function, minimizing $Q(R)$ is equivalent to maximizing $R$. According to the BER expressions of 2-PPM for SDSD and SDJD, i.e., \eqref{eqSDSD} and \eqref{eqSDJD-2}, we define a variable $R$ as
\begin{equation}
R={\frac{{\left( {{\cal X } + 1} \right)\gamma }}{{\sqrt {{{\cal X}^2}\gamma  + 2\gamma  + 2c} }}},
\end{equation}
where ${\cal X}$ is determined by the specific receiver. The derivative of $R$ with respect to ${\cal X }$ is given~by
\begin{equation}
\frac{\partial R}{\partial {\cal X }}=\frac{\gamma }{{\sqrt {\gamma {{\cal X }^2} + 2\gamma  + 2c} }} - \frac{{{\gamma ^2}{\cal X }({\cal X } + 1)}}{{{{\left( {\gamma {{\cal X }^2} + 2\gamma  + 2c} \right)}^{3/2}}}}.
\end{equation}
It can be calculated that
\begin{equation}
\left\{ {\begin{array}{*{20}{c}}
	{\frac{{\partial R}}{{\partial {\cal X } }} > 0,\qquad {\cal X}  \in \left[ {0,\frac{{2(\gamma  + c)}}{\gamma }} \right)},\\
	\frac{{\partial R}}{{\partial {\cal X } }} = 0,\qquad {\cal X}=\frac{{2(\gamma  + c)}}{\gamma },\qquad\\
	{\frac{{\partial R}}{{\partial {\cal X } }} < 0,\qquad {\cal X}  \in \left( {\frac{{2(\gamma  + c)}}{\gamma },1} \right]}.
	\end{array}} \right.
\end{equation}
Therefore, the maximum $R$ can be achieved at ${\cal X}=\frac{{2(\gamma  + c)}}{\gamma }$.

\begin{figure}[!t]
	\color{black}
	\vspace{-0.3cm}
	\centering{\includegraphics[width=3.6in]{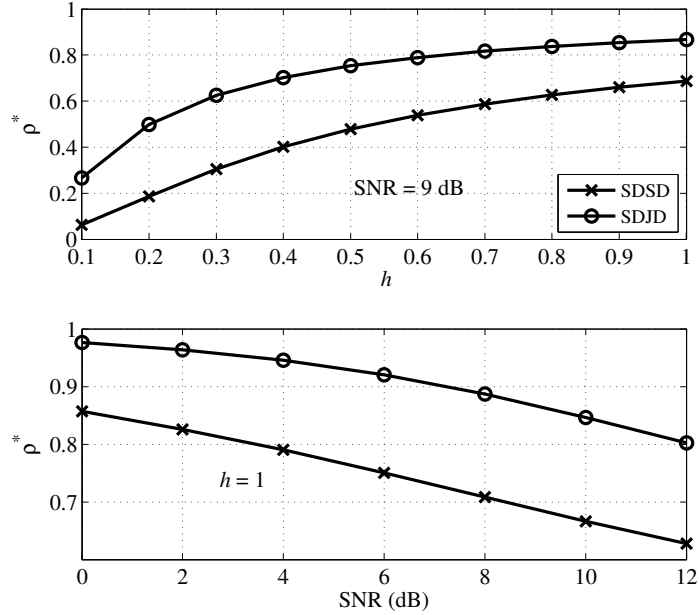}}
	\caption{Optimal splitting ratio $\rho^*$ versus (top) $h$ in ${\rm SNR}=9$ dB and versus (bottom) ${\rm SNR}$ in $h=1$. }
	\label{Functions}
	\vspace{-0.6cm}
\end{figure}

Specifically, according to \eqref{eqSDSD}, when ${\cal X}={\sqrt {\frac{\rho }{{{{\left( {1 - \rho } \right)}^2}{h^2}}}} }=\frac{{2(\gamma  + c)}}{\gamma }$, SDSD can achieve the minimum BER. The optimal splitting ratio is calculated as
\begin{equation}
\rho^*=\frac{{\gamma  + 4{{\bar h}^2}\gamma  + 4{{\bar h}^2}c - \sqrt \gamma  \sqrt {\gamma  + 8\gamma {{\bar h}^2} + 8{{\bar h}^2}c} }}{{4\left( {{{\bar h}^2}\gamma  + {{\bar h}^2}c} \right)}}.
\label{eqRhoSD}
\end{equation}
According to \eqref{eqSDJD-2}, let ${\cal X}={\sqrt {\frac{{{\rho ^2}}}{{{{\left( {1 - \rho } \right)}^4}{{\bar h}^4}}}} }=\frac{{2(\gamma  + c)}}{\gamma }$, we have the optimal splitting ratio for SDJD as
\begin{equation}
\rho^*=\frac{{\left( {{\gamma ^2} + 8{{\bar h}^2}{\gamma ^2} + 16{{\bar h}^2}\gamma c + 8{{\bar h}^2}c^2 - \gamma \sqrt {32{{\bar h}^2}\gamma c + 16{{\bar h}^2}c^2 + {\gamma ^2}(1 + 16{{\bar h}^2})} } \right)}}{{8{{\bar h}^2}{{(\gamma  + c)}^2}}}.
\label{eqRhoJD}
\end{equation}
Furthermore, the average optimal ratio is calculated by $E_{\bar h}\{\rho^*\}$.

\emph{Remark 5:} It can be found in Fig. \ref{Functions} that the optimal splitting ratio increases with the channel coefficient. When the channel is in a good state having a large value channel coefficient, $\rho^*$ will become large. In this case, more power of the received signal will be allocated to CD to achieve lower BER. In the deep fading scenarios having a small value channel coefficient, the receiver has to split more power to ED to combat the fading effect.

We can also find in Fig. \ref{Functions} that the optimal splitting ratio decreases with the ${\rm SNR}$; thus, more power should be allocated to ED in high SNR regimes. The reason is that by increasing the same SNR, the BER gain of ED is higher than that of CD.

{\color{black}\subsubsection{BER for PPM with spreading codes}
Since the existing IBC systems, including TH-UsWB \cite{UsWBrate} and DS-UsWB \cite{DS-USWB}, use spreading codes to improve SNR, we study the BER of the SDSD and SDJD receivers when using PPM with spreading codes.

The transmitter first spreads each bit using a pseudorandom spreading code and then transmits the spread signals using 2-PPM \cite{UsWBrate}. Assuming that the length of a spreading code is $N_s$, the received signal in Fig. \ref{SDSD-JD} becomes
\begin{equation}
r(t)=\sum_{j=0}^{\infty}\sum_{i=0}^{N_s-1}h\sqrt{E_p}p(t-(jN_s+i) T_c- d_{j,i} \delta)+n_{0}(t),
\label{eq42}
\end{equation}
where $d_{j,i}$ denotes the $j$th transmitted bit spread by the $i$th element of the spreading code, e.g., suppose the spreading code is ``10'', the transmitted bit of ``0'' will be spread as ``01''. Note that the spreading code is known at the receiver, and template signals are adjusted by the true transmitted signals.

Using soft-decision at the receiver, it can be inferred that the BER with an $N_s$ length spreading code, conditioned on channel coefficients, can be calculated by multiplying $N_s$ to both $\gamma$ and $c$ in \eqref{eqSDSD} and \eqref{eqSDJD-2} for SDSD and SDJD, respectively.}

\emph{Remark 6:} With perfect channel estimation, the BERs of both the SDSD and SDJD receivers decrease with $N_s$, which means that a greater value $N_s$ improves BER performance. It is worth noting that, compared to the $N_s$ length spreading code, increasing SNR $N_s$ times (e.g., increasing the transmission power $N_s$ times) can achieve better BER performance, which will be shown in Section \ref{Sect: Sim-BER}. The reason is that when $N_s$ is increased, the increased symbol duration also increases square-noise from ED.

\vspace{-0.3cm}
\section{Simulation Results} \label{Sect:Simulation}
In this section, we study the performance of the SDSD and SDJD receivers in terms of channel capacity and BER via Monte Carlo simulations. {\color{black}In all the simulations, the operating frequency is set to 2 MHz, which is within the frequency range of the medical ultrasound transducer (i.e., 1-20 MHz) \cite{FDA}. The maximum transmission power is set to $100\ \mu W$ (i.e., ${\rm SNR}=20$ dB) over a transducer area of 1 $cm^2$, which is below the maximum limit ($720\,mW/cm^2$) imposed by the FDA \cite{Experimental,FDA}. We carry out simulations under Gaussian channels (i.e., $|h|=1$) to show the effect of channel noise on the receivers. Further, we refer to the channel coefficient in \eqref{eqNaka} to study the impact of channel fading. Based on the existing experimental results of intra-body channel modeling, the parameters of the generalized Nakagami distribution in \eqref{eqNaka} are $z=0.59$, $\Omega=0.05$, and $s=1.12$ \cite{uSDN}.}

\begin{figure}[!t]
	\color{black}
	\centering{\includegraphics[width=3.6in]{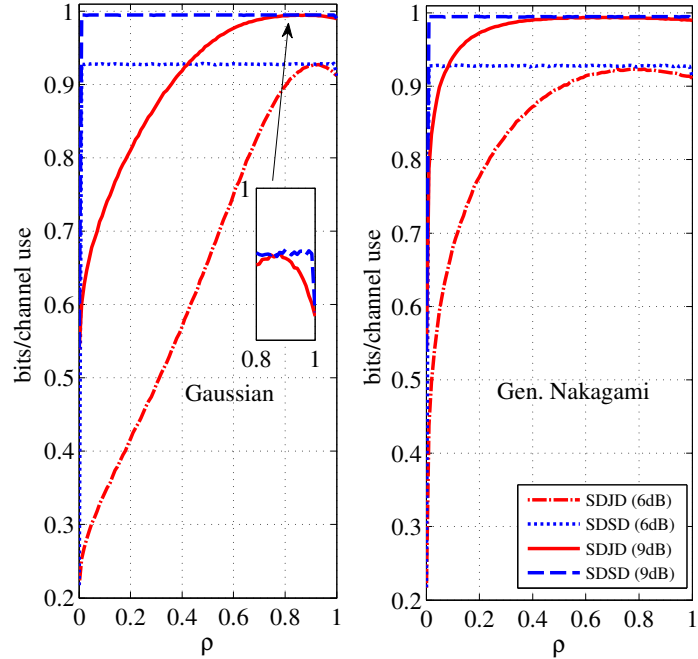}}
	\caption{Channel capacity of the SDSD and SDJD receivers in the case of 2-PPM and ${\rm SNR}=6,9$ dB over Gaussian and generalized Nakagami channels.}
	\label{Capacity}
	\vspace{-0.3cm}
\end{figure}

\subsection{Channel Capacity}
Figure \ref{Capacity} plots the channel capacity of the SDSD and SDJD receivers for the case of 2-PPM and ${\rm SNR}=6,9$ dB over Gaussian and generalized Nakagami channels. it can be observed that the SDSD receiver achieves stable channel capacity since the SNR in the independent CD and ED receivers is unaffected by the splitting ratio. The ED receiver (i.e., $\rho=0$) has the lowest channel capacity due to the severe square-noise from the non-coherent detection. The SDSD and SDJD receivers can achieve almost the same optimal channel capacity, which is higher than that of the CD receiver (i.e., $\rho=1$). The reason is that the joint use of CD and ED can increase channel capacity. Furthermore, we can observe that the SDJD receiver achieves its maximum channel capacity at $0<\rho<1$ due to the diversity combining. The optimal splitting ratio of SDJD becomes smaller as the SNR increases, and its value in the generalized Nakagami channel is smaller than that in the Gaussian channel at the same SNR. Thus, in the high SNR regime or fading channels, SDJD adopts a small $\rho$ to achieve the optimal channel capacity.

\begin{figure}[!t]
	\color{black}
	\centering{\includegraphics[width=3.6in]{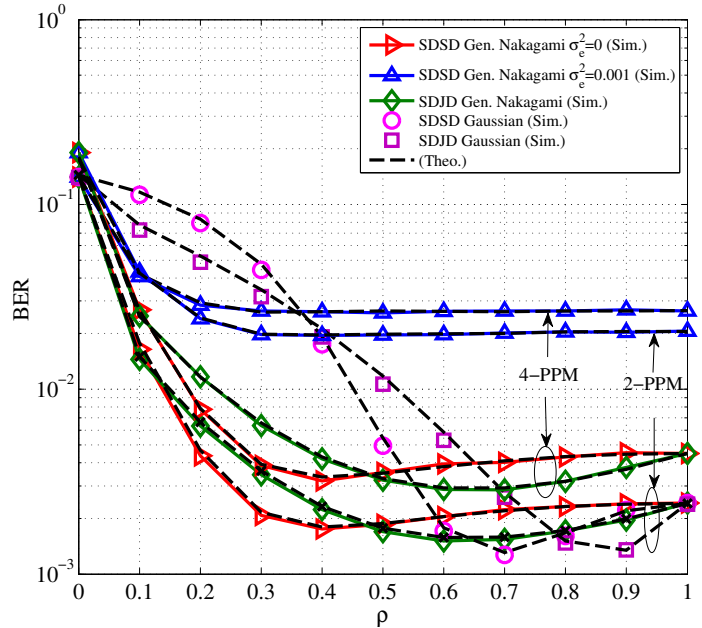}}
	\caption{Theoretical and simulation BER versus the splitting ratio $\rho$ using 2-PPM and 4-PPM, for the cases of $\sigma_e^2=0,0.001$, and ${\rm SNR}=9$ dB, over Gaussian and generalized~Nakagami channels.}
	\label{MPPM}
\end{figure}

\subsection{BER} \label{Sect: Sim-BER}
Figure \ref{MPPM} depicts the simulation and theoretical BER curves for the cases of 2-PPM and 4-PPM with ${\rm SNR}=9$ dB and $\sigma_e^2=0,0.001$, over Gaussian and generalized~Nakagami channels. It is clear that the theoretical curves agree with the simulation counterparts well, which validates the theoretical BER given in Section \ref{Sect: BER}. Although the CD receiver always achieves a higher channel capacity than the ED receiver as shown in Fig. \ref{Capacity}, the optimal splitting ratio for BER is neither $\rho=0$ nor $\rho=1$. The reason is that the corresponding CD and ED receivers do not exploit all the DoFs in the channel. Thus, the SDSD and SDJD (i.e., $0<\rho<1$) receivers can improve BER over the CD and ED receivers.

It can be also observed from Fig. \ref{MPPM} that the SDSD and SDJD receivers have different optimal ratios with which the BER can be minimized. This is due to the different decision criteria. In the generalized Nakagami channel, due to the fading effect, the optimal splitting ratio is smaller than that in the Gaussian channel, which is consistent with \emph{Remark 5}. As $\rho$ deviates from the optimal splitting ratio, the BER of SDSD and SDJD is close to that of the CD receiver. Thus, in the high $\rho$ region, error floors can be observed. More importantly, the channel estimation error, i.e., $\sigma_e^2=0.001$, deteriorates BER of the SDSD receiver greatly and result in much higher error floors than that of the SDJD receiver. Therefore, the SDJD receiver is more suitable for IBCs to combat the imperfect channel estimation.

\begin{figure}[!t]
	\color{black}
	\centering{\includegraphics[width=3.6in]{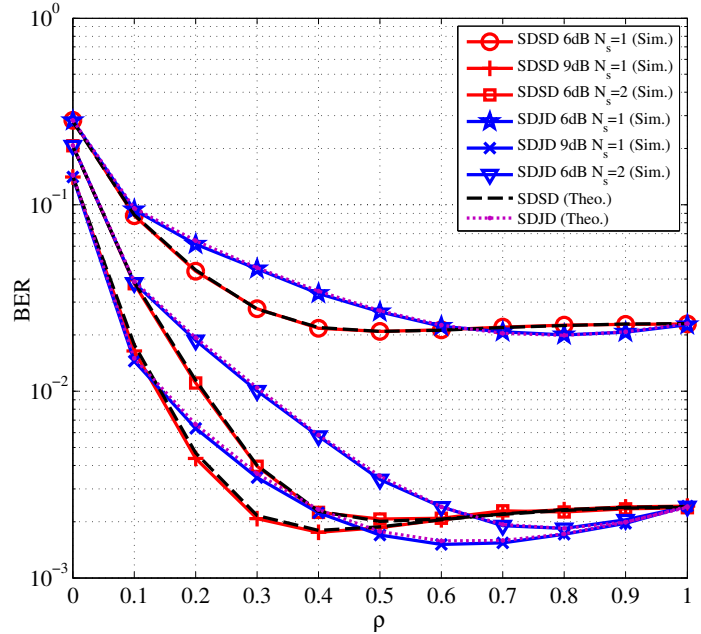}}
	\caption{Theoretical and simulation BER versus the splitting ratio $\rho$ using 2-PPM, for the cases of $N_s=1,2$, and ${\rm SNR}=6,9$~dB, over the generalized~Nakagami channels.}
	\label{SC-PPM}
\end{figure}

We also evaluate the performance of the SDSD and SDJD receivers under the transmission scheme of PPM with spreading codes, which is adopted in the existing TH-UsWB and DS-UsWB IBC systems \cite{UsWBrate,DS-USWB}.  Fig. \ref{SC-PPM} plots the theoretical and simulation BERs of the SDSD and SDJD receivers versus the splitting ratio $\rho$, for the cases of ${\rm SNR}=6,9$ dB,  when using PPM with spreading codes in the length of $N_s=1,2$, over the generalized~Nakagami channels. Comparing the minimum BERs of the SDSD and SDJD receivers with the CD receiver for the cases of ${\rm SNR}=6,9$ dB and $N_s=1$, the BER decreases by around  $0.4$ dB, $1.4$ dB and $0.6$ dB, $2$ dB for the SDSD and SDJD receivers, respectively. It means that as SNR increases, the minimum BER becomes much lower than that of the CD receiver, showing the advantages of the SDSD and SDJD receivers over the traditional CD receiver. Then, comparing the curves of ${\rm SNR}=6$ dB, $N_s=2$ with the curves of ${\rm SNR}=9$ dB, $N_s=1$. It can be observed that the CD receiver achieves the same BERs in the two cases, while the BER curves for the SDSD and SDJD receivers in the two cases are different. The reason is that ED introduces square-noise, while CD only has linear noise. Specifically, the minimum BERs of the SDSD and SDJD receivers in ${\rm SNR}=6$ dB, $N_s=2$ decrease by around $0.6$ dB and $1.2$ dB, respectively; while the minimum BERs in ${\rm SNR}=9$ dB, $N_s=1$ decrease by about $1.4$ dB and $2$ dB, respectively. Thus, increasing SNR $N_s$ times can achieve better BER performance compared with the $N_s$ length spreading code, which proves \emph{Remark 6}.

\begin{figure}[!t]
	\color{black}
	\centering{\includegraphics[width=3.6in]{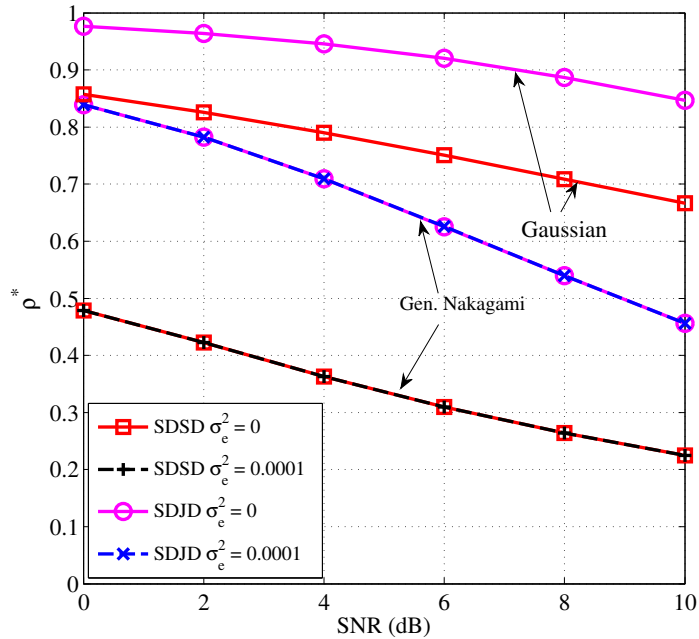}}
	\caption{Optimal splitting ratios $\rho^*$ of the SDSD and SDJD receivers versus SNR, for the cases of $\sigma_e^2=0,0.0001$, over Gaussian and generalized Nakagami channels.}
	\label{Ratio}
\end{figure}

\subsection{Optimal Splitting Ratio}
We then investigate the effect of channel fading on the optimal splitting ratio and the minimum BER of the SDSD and SDJD receivers. Fig. \ref{Ratio} depicts the optimal ratios of the SDSD and SDJD receivers over Gaussian and generalized Nakagami channels, for the cases of $\sigma_e^2=0,\rm 0.0001$. Obviously, $\rho^*$ in the Gaussian channels is higher than that in the generalized Nakagami channels, which means that more power is allocated for the CD receiver. The reason is that CD has a better BER performance than ED. As the SNR increases, ED has a higher BER gain than CD, and hence more power should be assigned to ED. We can also observe that the average $\rho^*$ is almost not affected by the small channel estimation error in generalized Nakagami channels. As SNR increases to $10$ dB, over half of the power in both the SDSD and SDJD receivers is allocated to ED to counteract the fading effect of the intra-body channels, due to the serious effect of fading on CD. Therefore, the optimal splitting ratio should be adjustable to accommodate different channel~conditions.

\begin{figure}[!t]
	\color{black}
	\centering{\includegraphics[width=3.6in]{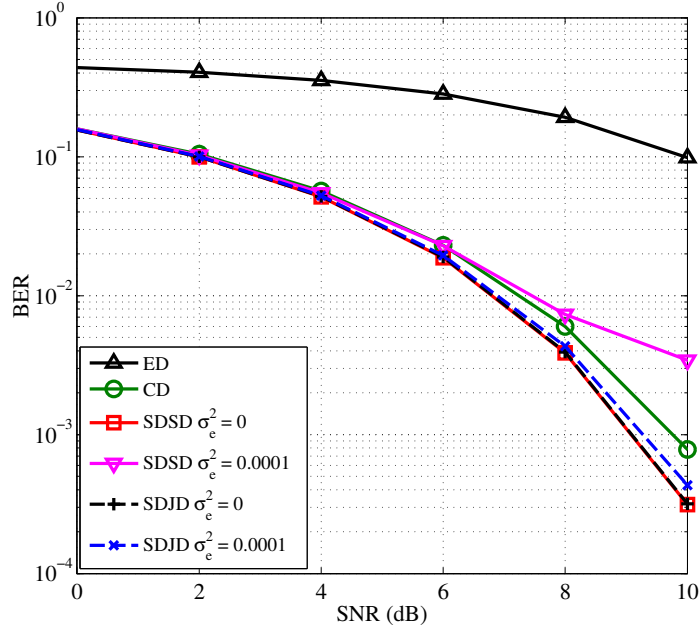}}
	\caption{Average BER versus SNR for the ED, CD, SDSD, and SDJD receivers, for the cases of $\sigma_e=0,0.0001$ over the generalized Nakagami channels.}
	\label{BER_ro}
\end{figure}

Using the $\rho^*$ in Fig. \ref{Ratio}, we further compare the minimum BERs of the SDSD and SDJD receivers. Fig. \ref{BER_ro} depicts the average BER versus SNR for the ED, CD, SDSD, and SDJD receivers, for the cases of $\sigma_e^2=0,0.0001$ over the generalized Nakagami channels. Comparing all the receivers, we find that by using CD and ED jointly (i.e., SDSD and SDJD), the BER can be lowered, and SDJD achieves the lowest BER in the case of $\sigma_e^2=0.0001$. Interestingly, although the SDSD and SDJD receivers have different $\rho^*$ in the case of $\sigma_e^2=0$, as shown in Fig.~\ref{Ratio}, the minimum BERs of the SDSD and SDJD receivers are almost the same. This is because both the SDSD and SDJD receivers jointly use CD and ED, and the decisions are based on the ML estimation, leading to the same optimal BERs. Another finding is that a slight channel estimation error will result in disastrous BER deterioration in SDSD, which coincides with the results in Fig. \ref{MPPM}. On the other hand, the SDJD receiver is slightly affected by the channel estimation error, based on the fact that the decision of SDJD does not use the estimated channel coefficient.

{\color{black}\section{Conclusions}}
This paper introduced a splitting structured receiver to ultrasonic pulse-based IBCs. An SDJD receiver was further proposed to simplify the complexity of the existing SDSD receiver and improve the BER performance in practical channels with channel estimation errors. The channel capacity, BER, and the optimal splitting ratio of SDSD and SDJD have been extensively analyzed. Our study showed that SDJD has substantial advantages over CD, ED, and SDSD receivers. The SDJD receiver can achieve the lowest BER comparing to other receivers with imperfect channel estimation. The complexity of SDJD is also lower than that of the SDSD receiver, which makes SDJD more suitable for the size- and energy-constrained intra-body biosensors. Our future work will focus on multiple implants with the SDJD receiver for intra-body networks, and the design of coding schemes to improve the BER performance of IBCs.

\vspace{0.3cm}
{\color{black}\section*{Acknowledgment}}

The authors would like to thank anonymous reviewers for their constructive comments that have significantly improved the technical content and presentation of this work.

\color{black}

\end{document}